\documentclass[%
 preprint,pre,
 amsmath, amssymb,
 aps,
 nofootinbib,jor,sor
]{revtex4-1}
\usepackage{enumitem}
\usepackage{graphicx}
\usepackage{dcolumn}
\usepackage{bm}
\usepackage{indentfirst}
\usepackage{caption}
\usepackage{subcaption}
\usepackage{amsmath}
\usepackage{amsfonts}
\usepackage{bm}
\usepackage{outlines}
\usepackage{xcolor,cancel}

\begin{document}

\preprint{APS/123-QED}

\title{Vibrational Effects on the Coefficient of Restitution}

\author{Satyanu Bhadra}
\email{satyanu.bhadra@tifr.res.in}
\affiliation {Department of Condensed Matter Physics and Materials Science, Tata Institute of Fundamental Research, Mumbai 400005, India}
\author{Shankar Ghosh}
\affiliation {Department of Condensed Matter Physics and Materials Science, Tata Institute of Fundamental Research, Mumbai 400005, India}


\begin{abstract}
A ball dropped from a given height onto a surface, will bounce repeatedly before coming to rest. A ball bouncing on a thick plate will behave very differently than a ball bouncing off the thin lid of a container. For a plate with a fixed thickness, a ball bouncing at the edge of a plate will be very different from the ball bouncing off the middle of the plate. 
We study the coefficient of restitution $\epsilon$ for a steel ball bouncing steel plates of various thicknesses. We observe how $\epsilon$ changes as the ball repeated bounces and finally comes to rest. Generally, $\epsilon < 1$ due to the dissipation of kinetic energy of the ball into the plate. However this dissipated energy can come back into ball in its later bounces. We see the emergence of super-elastic collisions ($\epsilon > 1$), implying that the ball gained Kinetic Energy due to the collision with the plate. We can increase the probability of such super-elastic collisions (P$_{SE}$) by adding a spring to the ball. We construct a simple theoretical model where the energy lost from previous collisions are transferred back into later ones. This model is able to simulate the occurrence of such super-elastic collisions.
\end{abstract}

\pacs{Valid PACS appear here}
\maketitle
\section{Introduction}
\rm{

A ball falling on a plate eventually comes to a stop. How exactly it comes to a stop is still not completely understood. The simplest approach to the problem is thus: At each collision the ball jumps to a height which is a fraction smaller than the height it fell from. The loss in the height accounts for the energy lost during collision. Since only a fraction is lost, the ball will bounce for infinite times till all the energy has been dissipated - all within a finite time. The process of infinite collisions in finite time leading to a halt is known as an inelastic collapse. Such inelastic collapse processes have been theoretically and numerically seen in 1,2 and 3 dimensions where the number of particles can vary from as low as 2 to infinitely many \cite{PhysRevE.87.042201,RevModPhys.71.435,PhysRevE.54.623,PhysRevE.50.R28,doi:10.1063/1.858323}. \par

However, our daily experience tells us that the ball stops bouncing after a few times. Dissipative and Adhesive mechanisms originating from the Van Der Waals interactions and Formation of Capillary Bridges between the surfaces of the plate and ball are responsible for bringing the ball to a halt. If the energy of the collision is less than that of the adhesion energy, the ball adheres to the plate and stops bouncing, i.e. it loses all its kinetic energy. This becomes prominent during the final bounces of drop heights of 50-100 nm, at which scale various adhesive and capillary attractive mechanisms, and the effect of small surface imperfections, become dominant \cite{delrio2005role,maboudian2004surface,israelachvili1972measurement,casimir1948influence,fuller1975effect,maugis1996contact}.\par

Considering the plate has been machined, we assume it has a surface roughness of around 100 nm. Within this rough surface, there will always be some material or impurity deposits, and capillary bridges can form easily. We assume these impurities have a surface tension of $\approx 10^{-2}$ N/mm.  For those scales Energy per unit area scales as $\approx 10^{-6},10^{-6},10^{-2}$ J/mm$^2$ for Casimir Interaction, Van Der Waals interaction, and Surface Tension (Capillary Bridges) respectively. The ball at the final velocities (10$^{-3}$ m/s) has a kinetic energy $\approx 10^{-8}$ J. If we consider an interaction area of 1mm$^2$ between the ball and the plate, we see that the surface tension energies are of the same order as the Kinetic Energy. \par

Thus the bouncing of a ball on any surface is inelastic, as the energy of the ball is being dissipated into the plate to overcome attractive forces, and lost to the surroundings, and the ball eventually stops. The inelasticity in the problem is captured by a single parameter, called coefficient of restitution, $\epsilon$ which is defined as 
\begin{equation}
  \epsilon=\left(\frac{h_{n+1}}{h_n}\right)^{1/2} =\frac{v_n^o}{v_n^i}
\end{equation}

Here, $h_n$ and $h_{n+1}$ are the maximum heights gained by the ball in the n and n+1 collision, while $v_n^0$ and $v_n^i$ are the velocities of the ball just prior and after the collision. \par
 This apparently simple problem has attracted sufficient interest over time. Hertz \cite{LandauElasticity} made the earliest study of the problem (1881-82). He assumed that as the ball hits a rigid surface it gets deformed, the kinetic energy of the ball gets stored as elastic energy in the ball. In time, the ball uses its elastic energy to bounce up. \par

\begin{figure}[tb]
     \includegraphics[width=.4\textwidth]{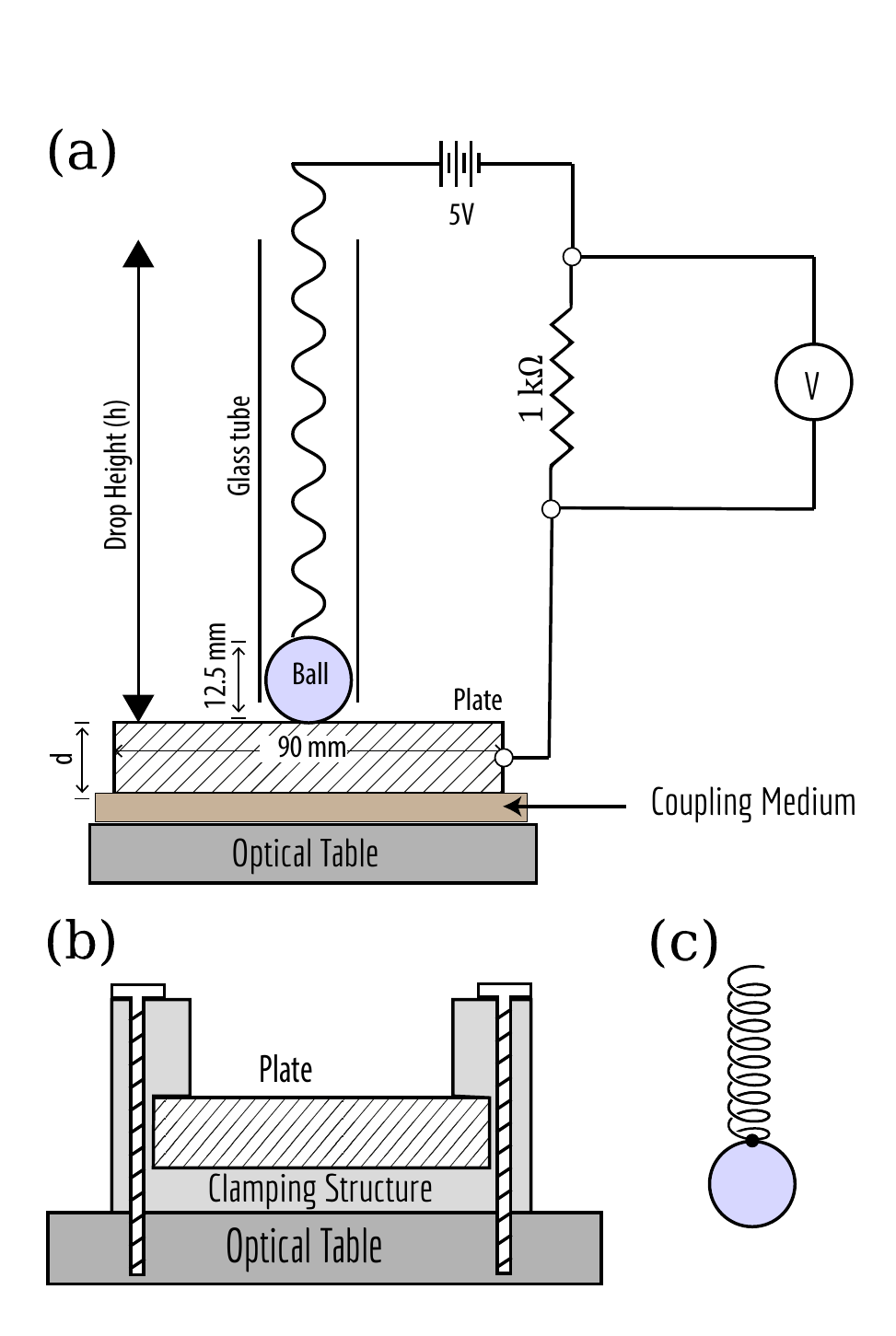}
   \caption{(a)The experimental setup with the ball and plate completing the electrical circuit. (b)The clamping structure. The plate was fixed securely within this structure, which was screwed to the optical table. (c) To add vibrations, a spring was welded on to the ball. }
    \label{fig:setup}
\end{figure}

 Hertz assumed that the duration of the collision was long enough to allow the complete energy transfer to occur, and all the energy from the collision would be confined to the local deformed surface. Thus in Hertz’s analysis, $\epsilon$ =1, i.e, no energy is lost in the conversion of the kinetic energy to elastic energy and back to kinetic energy. Raman (1920)\cite{RamanHS} published after much experimentation, how $\epsilon$ changed with the geometrical parameters like the plate thickness and introduced an dimensionless constant, the inelasticy parameter $\lambda$. Zener provided a theoretical justification of Raman’s result in 1941\cite{Zener1941} by  modifying the quasi-static approximation of Hertz. He was among the first to appreciate that the energy during the collision may not be localised, but transfers between the point of contact and the plate. In Zener’s model, $\epsilon$ was related to the amount of collisional energy absorbed by the vibrating and translational modes of the plate. The parameter $\lambda$ controls the amount of the collisional force that dissipates by moving the plate, given by the equation of motion
\begin{equation}
   \frac{d^2 s}{d t^2} + k s^{3/2} + \alpha \frac{dF}{dt} = 0
   \label{forceeq}
\end{equation}

Where $\lambda = \alpha \left(\frac{m^3}{k^{2} v} \right)^\frac{1}{5} $, k is a force constant, and s is the relative displacement between the plate and the ball, and F being the force due to the collision. \par
 If the target plate is massive and thick, and the velocity of the ball is sufficiently low, most of the energy of the collision is localised at the point of contact, and can come back to the ball within the time period of the collision, also called the time of contact (t$_C$). Thus at low impact velocities, $\epsilon \approx$ 1. Zener parameterized the geometrical and the mechanical aspects of the problem of collision of a ball (diameter $d_s$, Young’s modulus $E_s$ and Poisson’s ratio $\sigma_s$) with a semi-infinite plate (thickness $t_p$, Young’s modulus $E_p$ and Poisson’s ratio $\sigma_p$) into $\lambda$ and he showed that $\epsilon$ could be expressed in terms $\lambda$. In Zener’s model 
 \begin{equation}
  	\lambda=\frac{\pi^{3/5}}{3^{1/2}} \left(\frac{d_s}{2 t_p}\right)^2 \left(\frac{\rho_s}{\rho_p}\right)^{\frac{3}{5}} \left(\frac{v_o}{c}\right)^{\frac{1}{5}}\left(\frac{E_s^{\prime}}{E_s^{\prime} + E_p^{\prime}}\right)^{\frac{2}{5}}
  	\label{lambdaeq}
  	\end{equation}

\begin{figure*}[tb]
  \centering
  		\includegraphics[width=.45\textwidth]{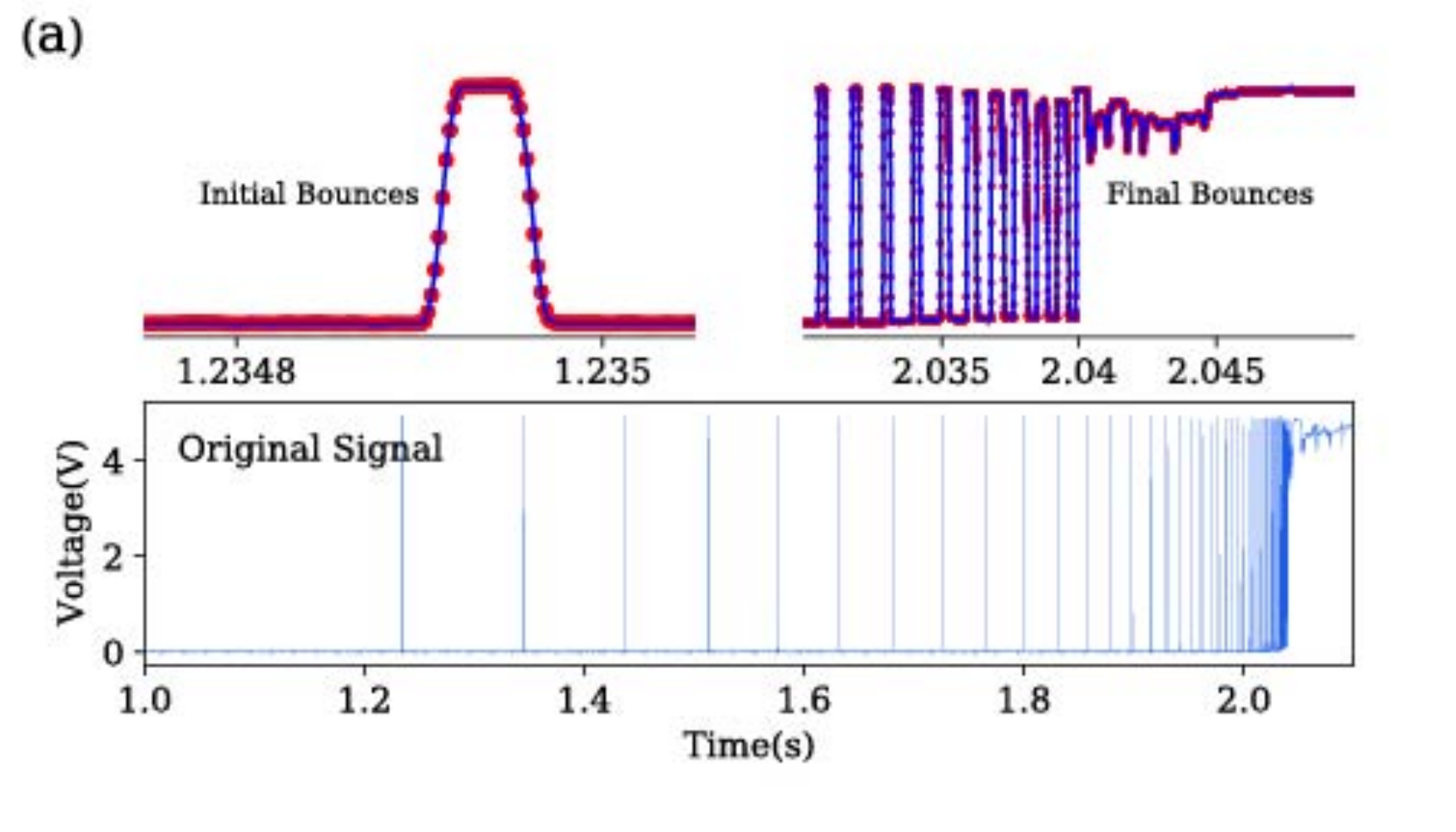}
  \includegraphics[width=.45\textwidth]{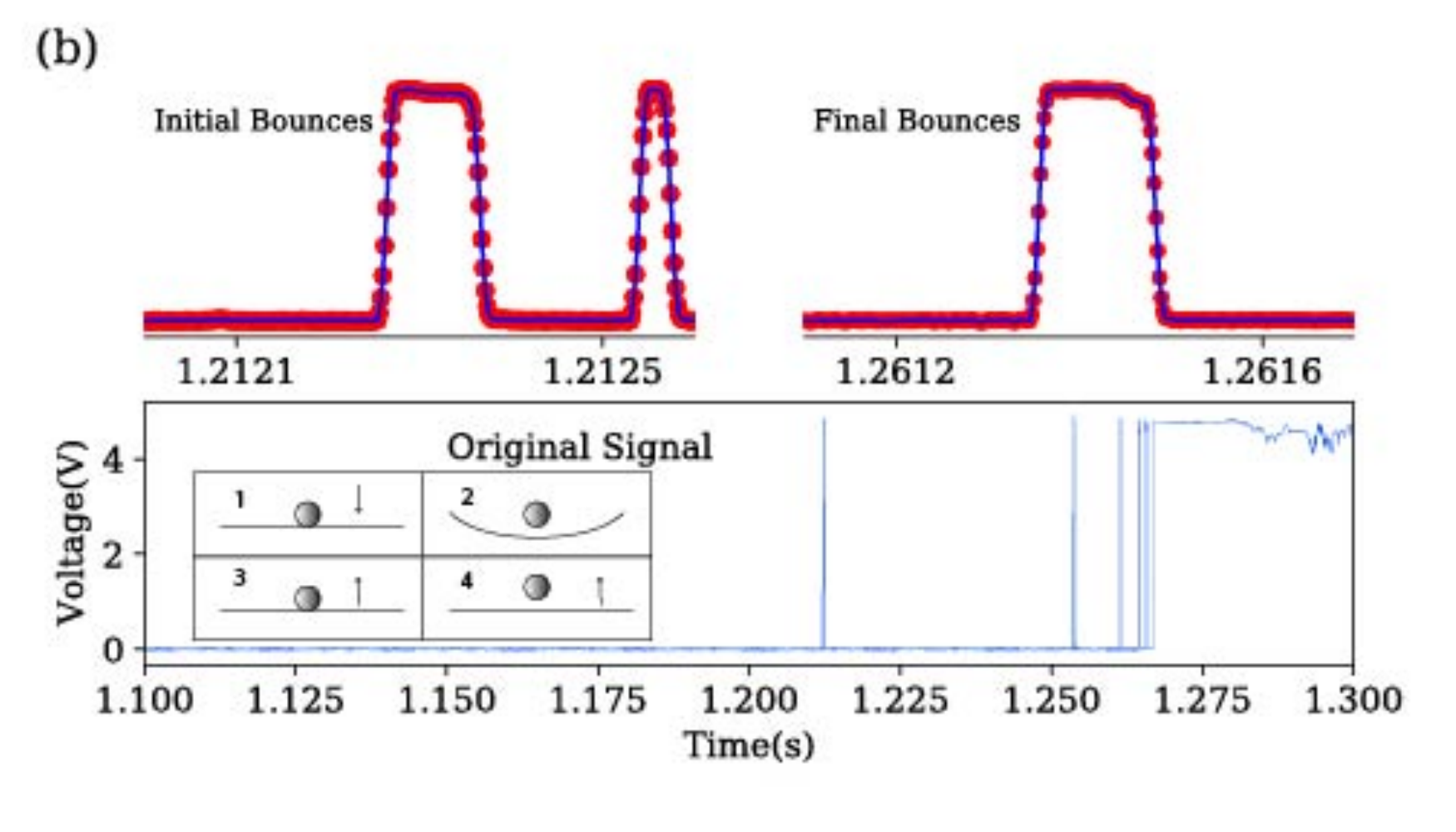}
  
  \caption{(a) Typical signal from most plates.  We have single clean peaks that become closely spaced as the ball comes to rest (High continuous voltage). (b) In the 2 mm plate however we have dual contacts, which eventually reduce to single peaks as the ball slows down. Inset : A Schematic of the process leading to the dual contact, consisting of 4 steps - $(1)$ Contact.  $(2)$Flexing of plate.  $(3)$Flexing back and second Contact. $(4)$Bouncing Back. }
  \label{Proper Contacts}
\end{figure*}

Here $c=\sqrt{\frac{E_p^{\prime}}{\rho_p}}$ is the velocity of sound waves in the plate and $E_s^{\prime} =\frac{E_s}{1-\sigma_s^2}$ and $E_p^{\prime}=\frac{E_p}{1-\sigma_p^2}$ and $v_o$ is the impact velocity. It is important to note that the kinetic energy of the system (ball and plate) remains constant in Zener’s model, the colliding ball imparts its kinetic energy into moving the plate, and during contact duration the ball does not receive its entire energy and hence leaves the plate with only a part of its energy. The rest of the energy is stored in the vibrational modes of the plate. However, in practice, there are lossy mechanism such as plastic deformations, and the kinetic energy of the system does not remain conserved \cite{Sondergaard1990}. Thus, Zener’s model is known to overestimate the value of $\epsilon$.  \par

 Additionally, the ball might lose energy via viscous dissipation due to deformation of the colliding surfaces.  The treatment of these losses is the same as Zener's model, with the $\lambda$ parameter being replaced by the Viscous loss parameter, A 
 \begin{equation}
   A=\frac{1}{3}\frac{(3 \eta_{2} - \eta_{2})^2 }{(3 \eta_{2} + 2 \eta_{2})}\frac{(1-\nu^{2})(1 - 2 \nu)}{E \nu^2}
 \end{equation}
 where $\eta_1$ and $\eta_2$ are the viscous constants that relate the deformation rate tensor and the dissipative stress tensor. \par
 
 In general, the following holds true for the coefficient of restitution, (i) its value is usually less than that of one, (ii) it depends on the velocity of impact. For small values of impact velocity, the plastic deformation of the materials in contact is minimal and $\epsilon$ is close to one. With increasing impact velocity, the plastic deformation increases and hence $\epsilon$ decreases \cite{ktgranular}. Thus, in the process of a bouncing ball coming to a rest, the impact velocity decreases and the coefficient of restitution increases with each bounce. \par

However these models involve the target being infinitely large - they are essentially semi infinite planes with a certain thickness. We challenge this notion by introducing using plates of finite size and thickness. In the following note, we revisit the problem of collision of a stainless steel ball ($d_s$=12.5 mm and weight 8.78 g) with a stainless steel disks or plates with varying thickness (diameter 90 mm and thickness $t_p$=2, 5, 10, 12.5, 20 mm). \par

In our experiments the ratio, $d_{s}/t_{p}<1$, moreover, the ratio of the diameter of the disk to that of the ball is about 7. The plate cannot be considered infinitely larger than the projectile. Our key findings are the following. 
\begin{itemize}
\item Vibrations set up in the plate due to repeated collisions affect how the ball bounces to a great degree. In the thinnest plates, the effects become more prominent – far reduced number of bounces, very lossy collisions, improper contacts during collisions(Fig \ref{Proper Contacts}). The effects can be controlled, and reduced by reducing vibrations in the system. \par
	
\item Super-elastic collisions ($\epsilon > 1$) occur at lower velocities. While this might seem to violate energy conservation laws, this is entirely possible if the energy dissipated into the plate does not die out instantly, but remains till the next bounce. The probability of super-elastic collisions can be greatly enhanced (Fig \ref{fig:zener}a,b) by adding a source of vibrations to the projectile i.e. by welding a spring on to the metal ball (Fig \ref{fig:setup}c) . 

\item We developed a simple physical model where the energy of the vibrations in the plate remain till the next bounce, and a fraction of that leftover energy may come back into the next bounce. We see that such a simplistic model is able to produce results which are quite satisfactory (Fig \ref{fig:Simulated}).
\end{itemize}

While a sizable amount of literature exists on how vibrating the target surface can affect the bouncing of the ball - the focus has been on driven systems, where the source of these vibrations are perfectly calibrated and set by the user to a specific frequency and amplitude \cite{chastaing2015dynamics,PhysRevE.48.3988,halev2018bouncing} .

}
\section{Details of the experiment}
\rm{

There are many methods as to how to collect experimental data of the value of $\epsilon$ in the regard of repeated collisions- High Speed Photography \cite{hsp2,hsp1,hsp3}, Pendulum Systems \cite{Pendulum_sys1,Pendulum_sys2,Pendulum_sys3}, LASER Doppler Velocimetry \cite{LDV},Particle Tracking Velocimetry \cite{PTV} amongst others. The most prominent technique is measuring the time lag between consecutive impacts \cite{KINGMENON, aUDIO}. However unlike previous experiments, we only measure the time lag from the electrical contact data and not the audio/piezo measurements, as they may lead to more sources of errors\cite{AUDIO}, especially as we are measuring in a very small target, in which vibrations persist longer than the usual massive plates used. \par

A schematic of our experimental setup is given in Fig \ref{fig:setup} (a). We carry out this study using stainless steel(grade 304) ball (diameter 12.6 mm, weight 8.78 g) and plates of various thicknesses ranging from 20 mm thickness to 2 mm thickness, each having a diameter of 90 mm. The voltage is supplied and measured using by a NI BNC 2090A and a NI M series 6259 respectively, with the ball and plate acting like a switch. We obtained the data at a rate of 500kHz, or with a time resolution of 2 $\mu$s.  \linebreak
In our experiments, we define
\begin{equation}
  \epsilon_i = \frac{v_{i+1}}{v_i}=\frac{t_{i+1}}{t_i}
\end{equation} 
where $t_i$ is the time gap between the $i^{th}$ and $i+1^{th}$ collisions. We can measure the time of contact, and width of the signal, and the time delay between peaks to great accuracy.

\begin{figure*}[t]

\includegraphics[width=.44\textwidth]{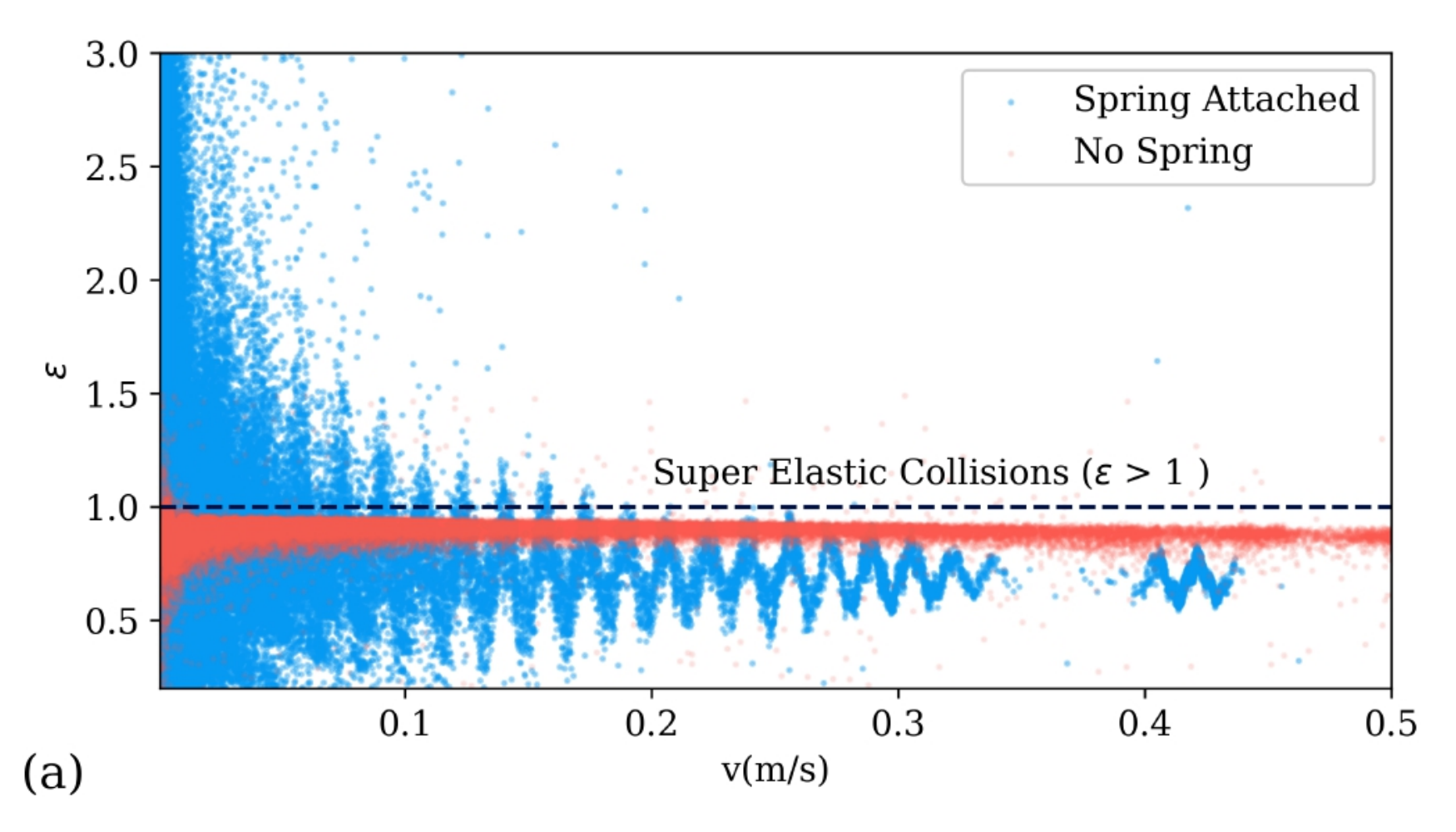}
\includegraphics[width=.43 \textwidth]{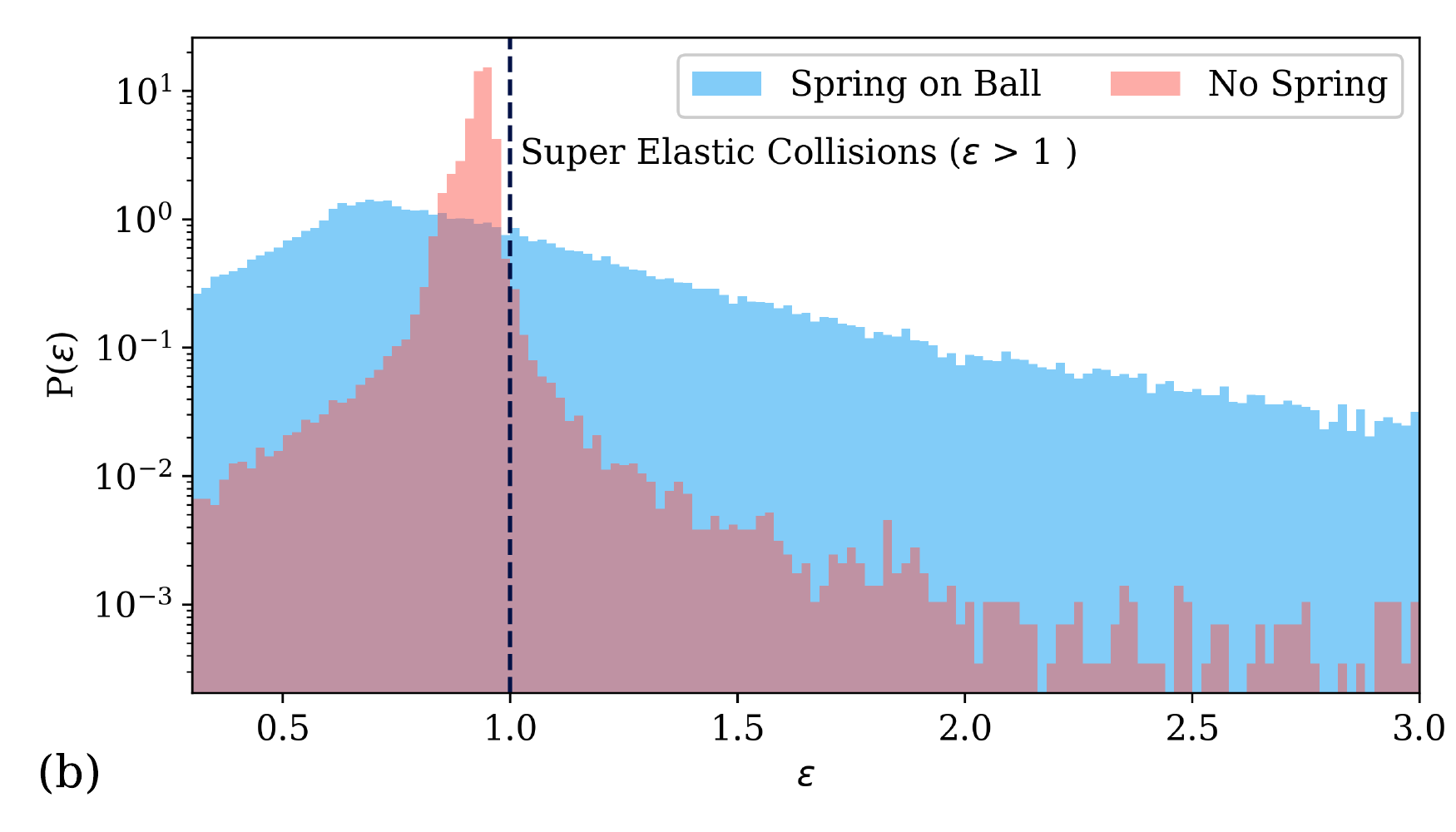}

\caption{(a)Adding a spring leads to highly regular peak like structures, and huge variations in the $\epsilon - v$ data. (b) The Probability distribution $P(\epsilon)$ flattens and spreads due to the spring.  }
\label{fig:zener}
\end{figure*}

 To limit the effect of vibrations, we have built a clamping structure (Fig \ref{fig:setup}(b)). This prevents larger flexing modes of the plate, enforcing the condition that the plate is stationary at the boundaries. The ball is raised to a height of 6cm above the target, and dropped by an Arduino based robotic arm, ensuring the ball hits the target at nearly the same velocity and location at each drop, and the experiment could be repeated thousands of times without any human intervention. On an average, nearly $10^4$ data points were generated per plate. We claim that the ball has stopped bouncing once its $\epsilon$ value reaches a value greater than 10 (signifying that it has come to rest). For all experiments, the plate was either placed directly on the optical table(on top of a 100 $\mu$m acetate sheet to provide electrical insulation), or in the clamping structure, that was screwed to the optical table. To study the effect of methods of coupling, we placed different materials (foam, paper) in between the plate and the optical table surface. To study the effect of added vibrations to the system, we added springs of varying stiffness to the steel ball via welding, to make a seamless contact (Fig \ref{fig:setup}c). 
 
\begin{figure*}[t]

\includegraphics[width=.45 \textwidth]{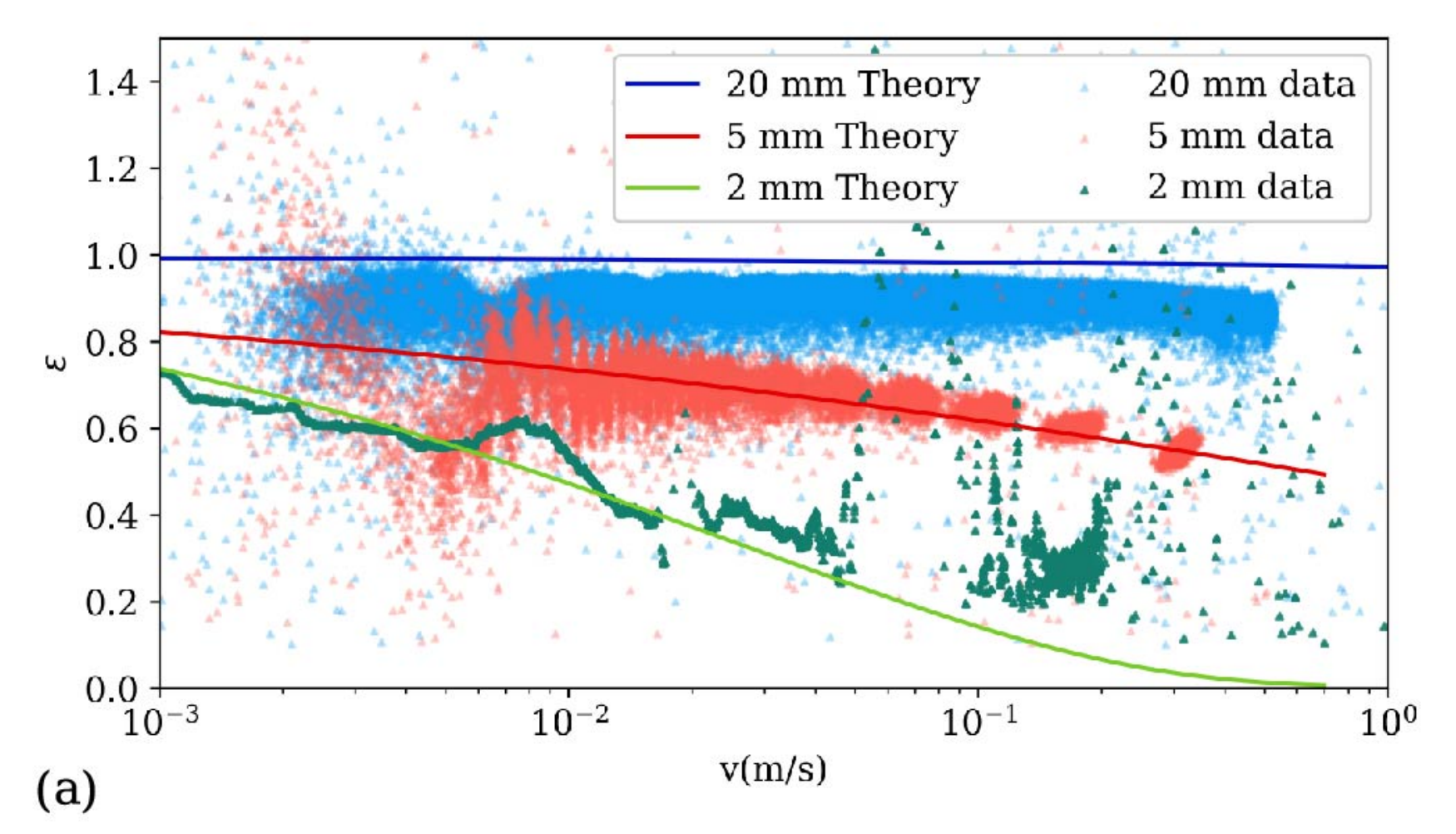}
\includegraphics[width=.45 \textwidth]{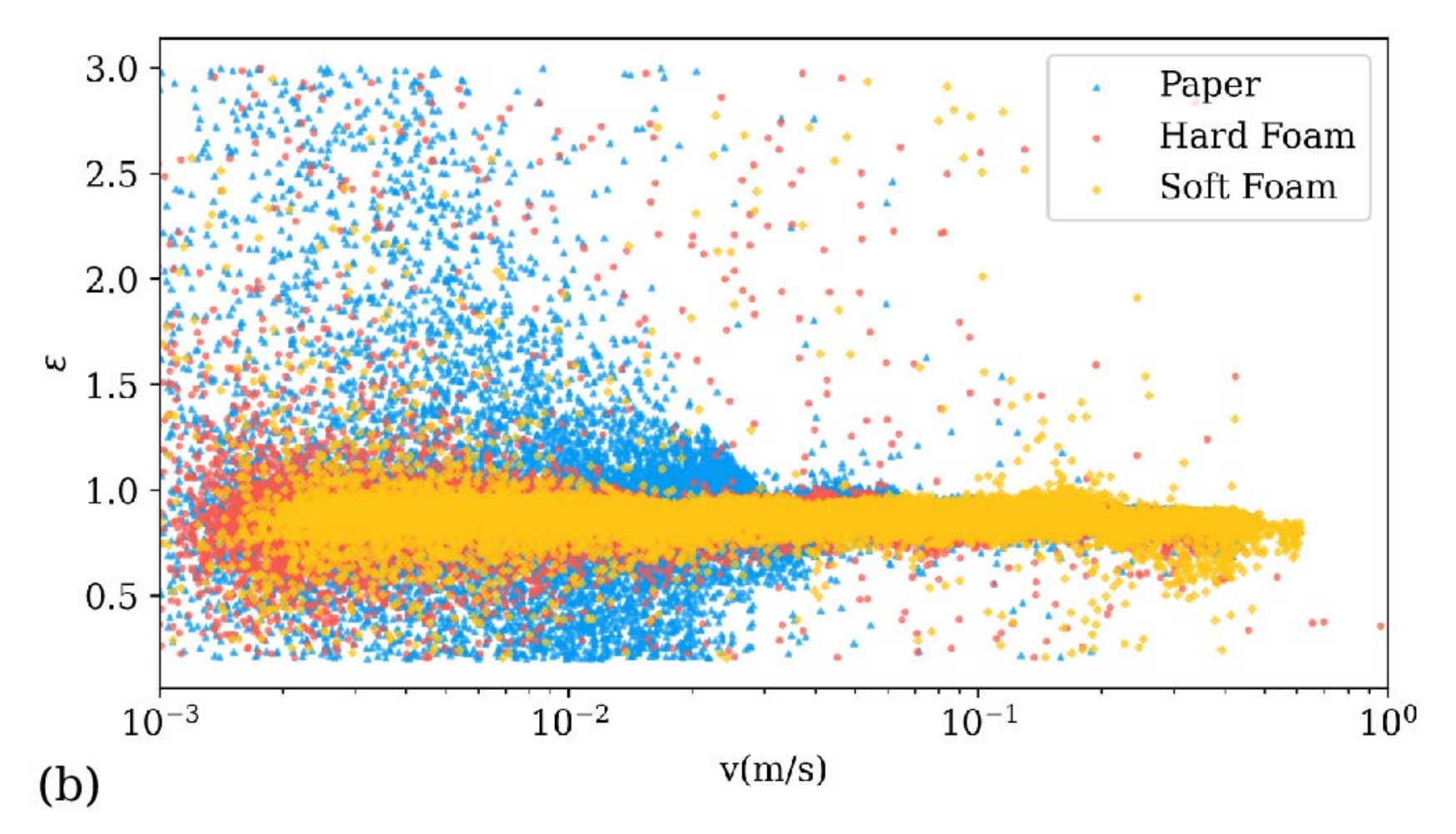}

\caption{(a) Deviation of experimental data from Zener's Model for 20mm, 5mm, 2mm plates. X axis is plotted in logscale. (b) On adding coupling media in between the optical table and the 20 mm plate, we see that the foam has hardly any effect. However when the stack of papers is used, we see very noisy data, below 0.03 m/s. }
\label{fig:scatter}
\end{figure*}

}

\section{Observations}
\rm{

To obtain Zener's Theoretical Values, we obtain $\lambda$ as a function, f(v,d) from Eq. \ref{lambdaeq}. We numerically solve Eq. \ref{forceeq} to obtain F(t) and time duration of the collision(t$_{C}$). We use the formula provided by Zener \cite{Zener1941} to obtain 
\begin{equation}
 \epsilon = \frac{1-R}{1+R}  
\end{equation}
where R = 2 m $\alpha \left(\int_{0}^{t_C} F^2 dt/(\int_{0}^{t_C} F dt)^2\right) $.  \par
Although the target is of a limited size we still see striking resemblance to results and experimental data obtained from experiments involving bulkier targets. We see the presence of visco-elastic losses at higher initial velocities across all plates, and observed values are lesser than that calculated by Zener.In the thinner plates, however, the energy being dumped into the plate does not completely decay away between collisions and comes back to the next collision due to the vibrations present. Thus we observe $\epsilon$ values being equal or even greater than expectation (Fig \ref{fig:scatter} a). As long as the coupling media (Fig \ref{fig:setup} a) between the plate and the optical table is a homogenous one, we see no effects on the $\epsilon$-v scatter , but if we use a layered media such as the stack of papers, we see very noisy data at lower velocities (v<0.03 m/s) (see Fig \ref{fig:scatter}b) \par
In the 2 mm plates, we see that when the ball makes contact, we have two closely-spaced peaks in the signal as opposed to one clean peak we got in the other plates (Fig \ref{Proper Contacts}). We call such contacts dual contacts or improper contacts. A significant amount of energy is dumped into the plate, and not recovered by the ball. Most collisions are extremely lossy and noisy, and bring the ball to an immediate stop. On placing the plate in the clamping structure, the probability of dual contacts drops to half. A schematic of the process is described below (Fig \ref{Proper Contacts} b, inset) 
\begin{itemize}
\item The ball falls on the plate and creates a contact, and causes it to flex. This gives us the first peak. 
\item The plate flexes downwards, and away from the ball. Meanwhile the ball has come to near rest, and thus stops being in contact. This gives us the short contact gap. \footnote{In the time between the two peaks as in Figure \ref{Proper Contacts}, the ball would have moved around 100 nm due to gravity, whereas the amplitude of the fundamental mode of the oscillation $\approx$ 10 $\mu$m with a frequency $\approx$ 10 kHz.}
\item The plate flexes back, contact is reestablished. This gives us the second peak.
\item The plate's flexing aids the ball in its bounce, and bounces up again.
\end{itemize}

As the ball comes to rest, the second peak nearly disappears, before we see that the peaks have merged to one. As the velocity drops, the ability of the ball to impart enough energy to have the large amplitude of oscillation required for loss of contact, reduces.

In the aforementioned models (Zener, Viscoelastic), the boundary conditions are such that the target is absolutely stationary at the moment of impact, the plate is held securely at the edges such that there is no motion there, and whatever energy is dissipated into the plate from the collision is lost immediately so that no information of it remains during the next collision. However in a finite system such as ours this cannot always be true. Therefore we see unexpected collisions where $\epsilon>1$ (Fig \ref{fig:zener},\ref{fig:scatter}). Such 'Super-elastic' collisions are extremely rare, but occur at lower velocities. Most occur when v $\in [0.001 m/s,0.01 m/s]$.  With a rise in Frequency of Super-elastic collisions there is also a fall in the value of mean $\epsilon$. Each plate has a resonant frequency ($f_{res}= t_{res}^{-1}$). We define $v_{res} =g t_{res}/2$. We see pronounced dips of $\epsilon$ at even multiples of $v_{res}$, with the extent of the dip increasing as one comes close to $v_{res}$.  Most of the collisions terminate at $v_{res}$, thus at that velocity, the ball is able to dump most of its energy into the plate. 
\par
As the plate gets thinner and reduces in mass, the kinetic energy lost from the ball sets up vibrations with higher amplitudes. Due to the lower mass and thickness, these vibrations take longer to damp out. The probability of super-elastic collisions for a given plate is given by
\begin{equation}
P_{SE} = \frac{\text{Total number of super-elastic collisions}}{\text{Total number of collisions}}  
\end{equation}
 P$_{SE}$ increases with the plate becoming thinner. With the addition of the clamping structure, vibrations are further damped, restricting the mechanism via which super-elastic collisions take place. As expected, P$_{SE}$ reduces if the plate is clamped. The ratio(R$_{fc}$) of P$_{SE}$ for free plates to P$_{SE}$ for clamped plates increases as the plate gets thinner(Fig \ref{fig:Ratio}). 
\par
Contact time or Time-period of collision ($t_C$) was relatively unchanged across all plates, clamped or free.(Fig \ref{toc} a). We placed the 20 mm plate over a intermediate medium between the plate and the optical table, to see if the material properties of the coupling medium would have any effects on the collision time. Naively we would expect that a softer medium would move with the plate, and hence increase contact times. We used soft and hard packaging foam, and a stack of A4 (300 pages, 75 gsm). Majority of the data lay in the same band as of the plates. In the stack of papers, we found band like structures (Fig \ref{toc} b), absent in all other cases. Due to the lack of cohesion in between individual sheets/layers, energy is dissipated via local shearing. In our systems, the time of contact diverges as velocity is reduced. This contradicts an earlier study, which claims that it should logarithmically tend to the Hertzian value. This assumption is based on the quasi-static process \cite{tocvsv}, does not take into consideration the leftover energies from the previous bounces and considers that the plate is absolutely still when the collision takes place.

\begin{figure}[bt]
  
   \includegraphics[width=1\linewidth]{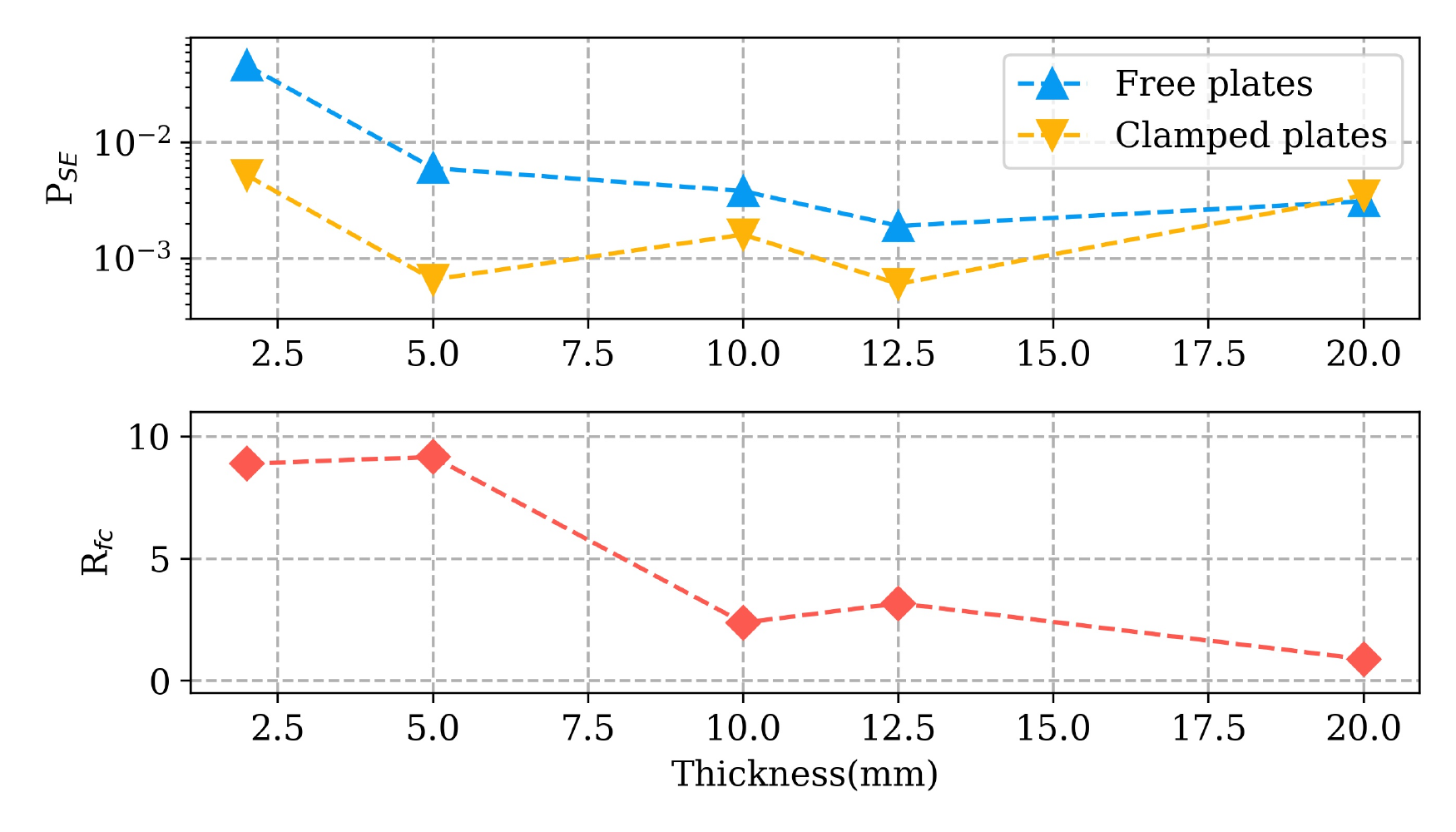}\hfill
  
  \caption{The effect of clamping is felt most clearly in the superellastic collisions. As the plate reduces in thickness, probablity in super-elastic collisions ($P_{SE}$) increases if kept free, but remains relatively unchanged if the plate is kept clamped (Note that y axis is in logscale). Hence The Ratio (R$_{fc}$) of the values rises with decrease in plate thickness. }
  \label{fig:Ratio}
\end{figure}

\begin{figure}[tb]

     \includegraphics[width=.45\textwidth]{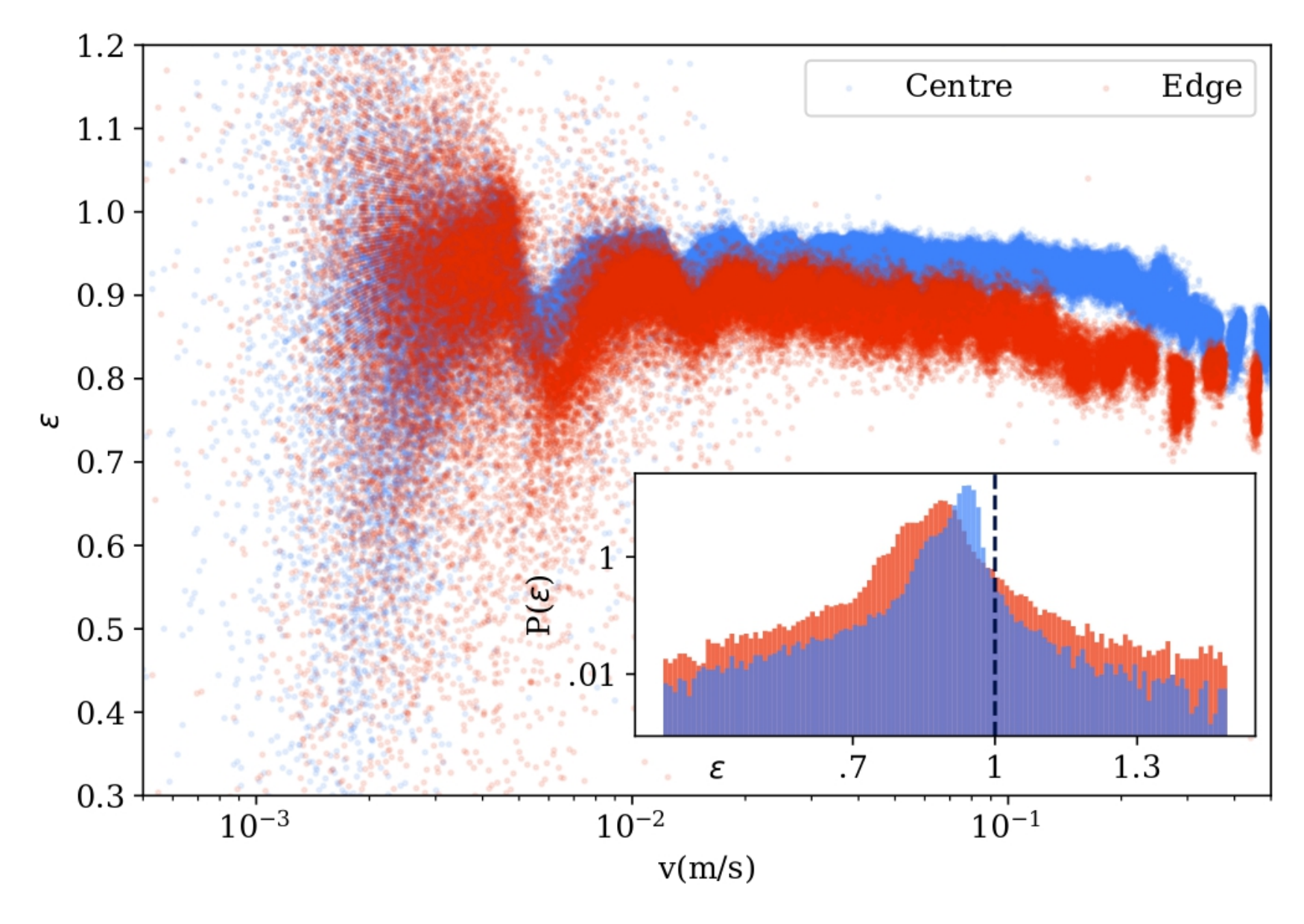}

   \caption{The collisions at higher velocities are more lossy (lower values $\epsilon$) at the edge(red) than at the centre(blue). As the ball slows down, most of the data overlaps. Inset : P$_{SE}$ is more pronounced at the edge, giving us an insight as to the role of vibrations in their genesis.}
    \label{fig:location}
\end{figure}

\begin{figure*}[tb]

     \includegraphics[width=.45\textwidth]{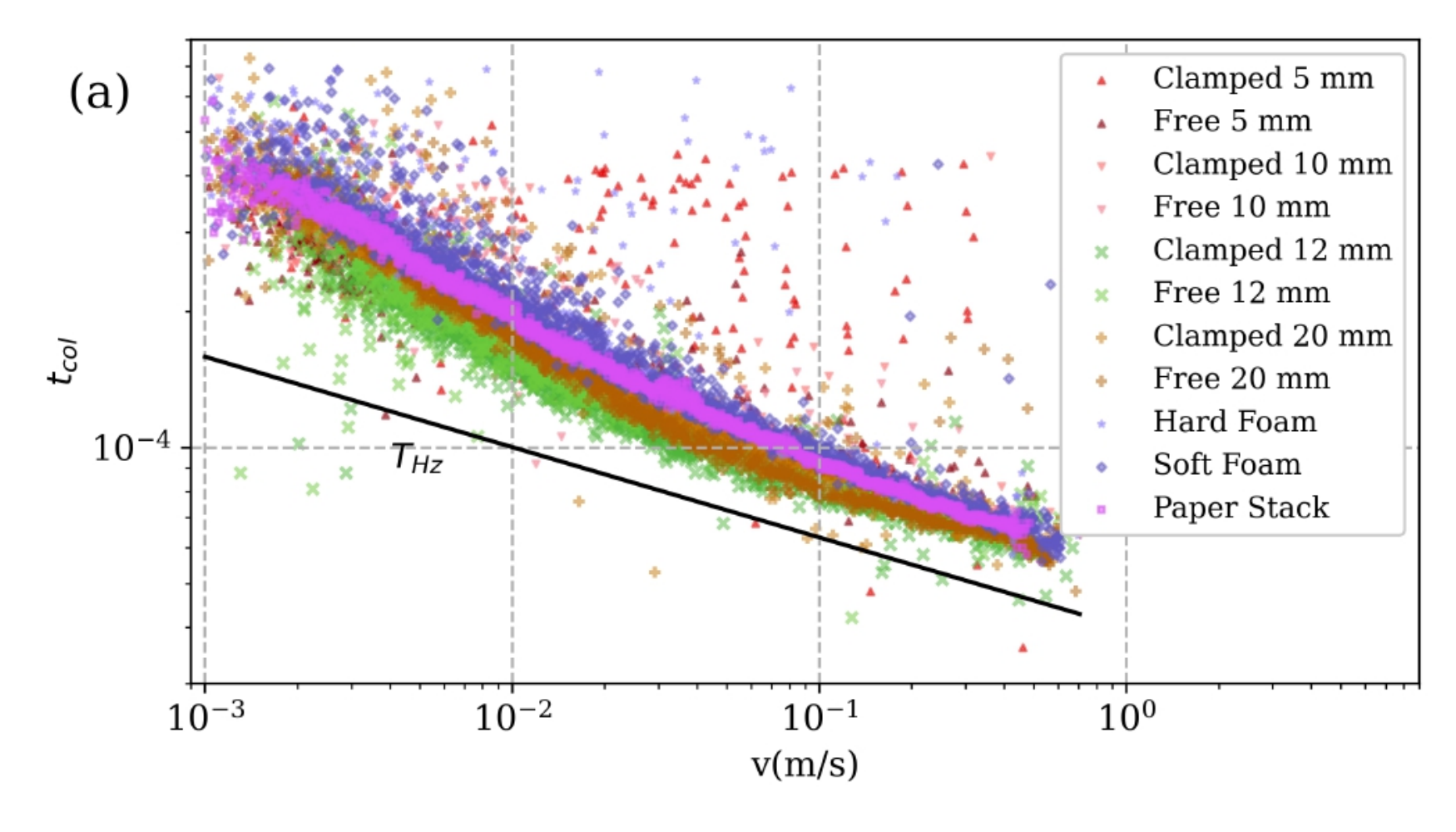}
     \includegraphics[width=.45\textwidth]{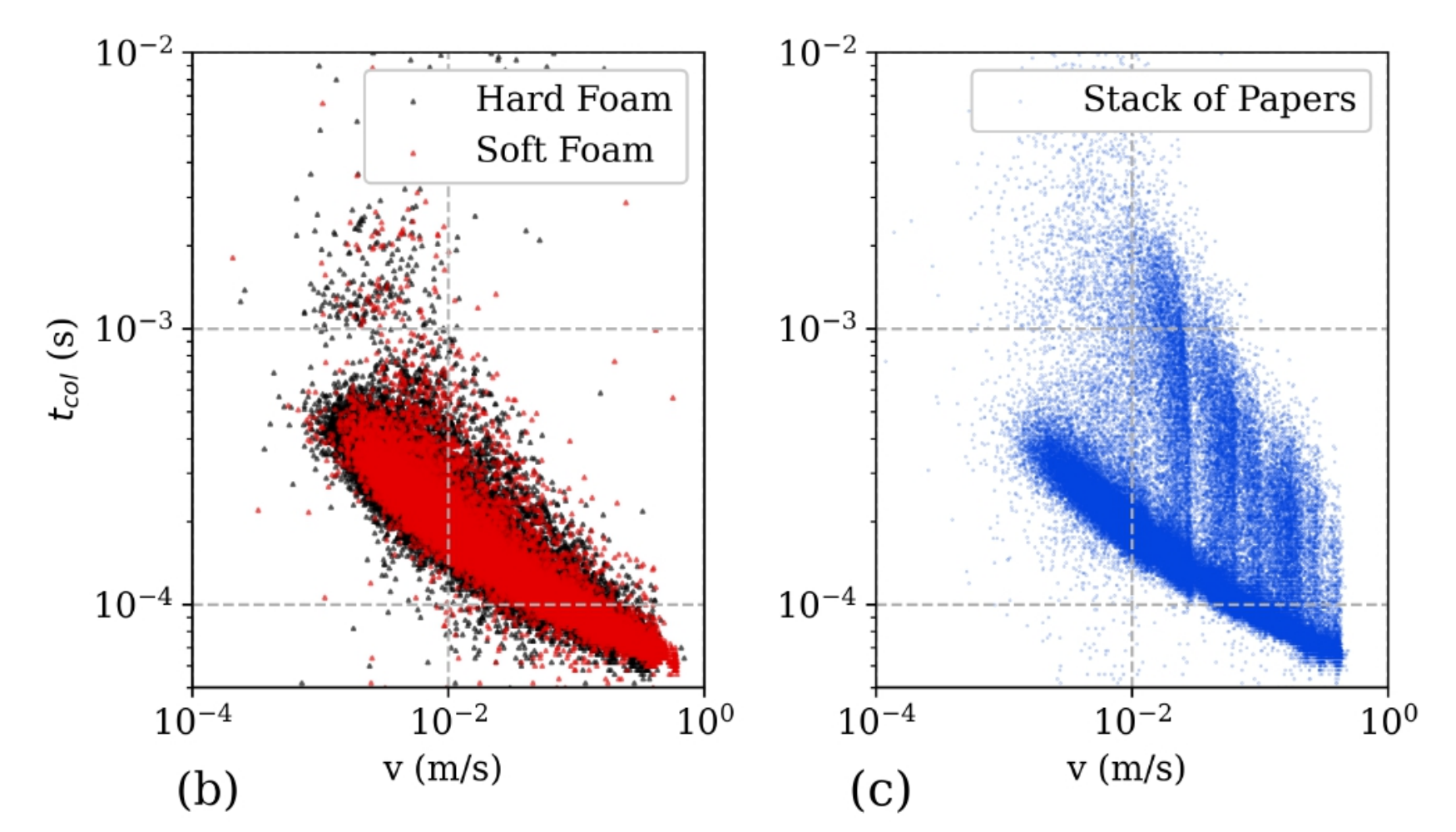}

   \caption{(a)The Average Contact time ($t_C$) is relatively unchanged over plates and changing coupling materials. Average values are calculated over logarithmically spaced bins in the velocity range. Instead of slowly converging to the Hertzian Values at lower Velocities, $t_C$ veers away from it. (b)The foam materials, as coupling media, do not have significant effect on the contact times.  However in the case of the stack of papers, we find that apart from the narrow region, we have the appearance of band like structures. This might be due to shearing between the individual sheets of the stack, making the process more sticky, thus having longer contact times.}
    \label{toc}
\end{figure*}
\par
The location of the ball bouncing on the plate is also an important parameter. We naively expect that as we apply our force further away from our centre of mass of the plates, there will be additional torque on the system, as well as edge effects.  impact the ball on our 20mm plate at Centre(within 0 to 5mm of the centre of the plate), and Edge (40 to 45mm). As expected, at the edge, we have fewer number of bounces than at the centre (40 vs 73), higher P$_{SE}$ (0.048 vs 0.018) and generally lower values of $\epsilon$ (Fig \ref{fig:location}).
\par
So far we were only interested in how the vibrations of the target plate were affecting the bounces. What if the projectile itself could add vibrations? We welded a spring to the end of the ball, and bounced the composite structure on the ball end (Fig \ref{fig:setup}c). By adding a spring to the bouncing ball, we have effectively added a new mode that can store or lose energy. While the centre of mass of the ball-spring system still follows a simple motion due to gravity, the two ends of the system has added motion due to the oscillations of the spring length.
\par

We see the emergence of oscillatory bands in the $\epsilon$-v scatter (Fig \ref{fig:zener} a). We performed the experiment with three different springs of varying spring constant(k(S1)=1744.8, k(s2) =1433.1, k(s3)=1169.3 N/mm respectively), all three of which show bands. As k decreases, the softer spring oscillates more easily. More energy is lost via the spring than the plate, and average $\epsilon$ decreases (showing less bounces). These bands have peaks which are separated by almost constant values. These inter-peak or peak-to-peak gaps($\Delta_{v}$) are nearly constant for each spring and decrease as the spring constant increases (Table \ref{springk}). 
\par
The most observable difference is the large variation in the distribution of $\epsilon$, and also a rise in P$_{SE}$ from low as. 02 to as high as 0.4 (Fig \ref{fig:zener} b). The value of $\epsilon$ is consitently observed as high as 3, compared to 1.2 for a non-spring system. However we see that the peak of this distribution has shifted from. 95 to. 65 on addition of a spring. This explains why the number of bounces observed is much lower than that of an unmodified or control ball.
\par
\begin{center}
\begin{table}
 \begin{tabular}{|c |c| c| c|} 
 \hline
 K(N/mm) & Mean Gap ($\Delta_{v}$)(m/s) & Average $\epsilon$ & P$_{SE}$ \\ 
 \hline
 1744.8 & 0.0165 & 1.014 &. 39 \\ 
 \hline
 1433.1 & 0.017 &. 9749 &. 36 \\
 \hline
 1169.3 & 0.0212 &. 9265 &. 30 \\
 \hline
 \end{tabular}
 \caption{We find that most observable results from our data correspond to the spring's stiffness. }
 \label{springk}
\end{table}
\end{center}
}

\section{Theoretical Model}
\rm{ We modify Zener's model to fit our observations. We assume that at each collision only a fraction ($\beta$) of the collision's momentum ($P_{C}$) is able to transfer into the ball, the rest is dumped into the plate and stored. $P_{C}$ is calculated from Zener's Model. The rest of the energy is dumped into the plate. This increases the momentum of the plate and this momentum of the plate is expressed as $P_{S}$.  As the plate vibrates it loses its stored energy(or momentum) due to dampening. We do not expect the stored momentum to be such that it is more than the momentum of the collision itself. So we introduce a lossy term that linearly grows with $\frac{P_{S}}{P_{C}}$.  During the next impact, a fraction $\gamma$ of the stored momentum goes into the collision, along with a random phase factor $\phi$.  As the stored momentum increases it can be lost in large amounts as flexures in the plate. The algorithm for each collision is now 
\begin{itemize}
  \item [1]. $(1+\epsilon) P_{I} = \beta P_{C} + \gamma Re(e^{i \phi})P_{S}$
\item [2]. $P_{S} = (1-\gamma) P_{S} + (1-\beta) P_{C}$
\item [3]. $P_{S} = P_{S} \times exp(-\frac{t}{\tau})\left(1-\frac{P_{S}}{P_{C}}\right)$
\end{itemize}

If we ignore the memory term $\gamma$, we find that $\beta_{exp}$ ($\beta_{exp} =\beta $ when $\gamma$ = 0) relates the experimental values of $\epsilon$ ($\epsilon_{exp}$) and the expected values from Zener's Model ($\epsilon_{Z}$) by the following equation.
\begin{equation}
  \beta_{exp} = \frac{1+\epsilon_{exp}}{1+\epsilon_{Z}}
  \label{betaeq}
\end{equation}
Using formula (\ref{betaeq}), we obtain how $\beta_{exp}$ changes with velocity. We choose $\beta$ to be a function of velocity by fitting it to $\beta_{exp}$-v data. We fit the initial portion at higher velocities to the standard power law series we obtain when dealing with Zener's model or Viscoelastic models\cite{ktgranular}. The fall in the values of epsilon at lower velocities from around 0.01 m/s is modelled as an exponential function. At velocities $\approx v_{res}$, the fluctuation about the average values is considered to be harmonic series. Given that most collisions terminate at v$\approx v_{res}$, we take $\beta = Random[0,1]$ at $v_{res}$. So $\beta$ has the form,
\begin{equation}
\begin{split}
\beta = & (A_{0} + A_{1} v^{1/5} + A_{2} v^{2/5} +. ..)\\
& \times (1 - B exp (-v/0.01)) \\
& \times \left (1 + \frac{C}{v^2} exp(i \pi \frac{v}{v_{res}})\right)
\end{split}
\end{equation}
where A$_{i}$, B, C are obtained from fitting $\beta$ to $\beta_{exp}$. \linebreak

With such a model for $\beta$, and $\gamma, \phi $ being uniformly random $\in$ [0,1], we are able to obtain close similarity to the experimental data (Fig \ref{fig:Simulated}). Thus accounting for the stored energy not being dissipated as easily as we see in bulkier plates, we are able to reproduce new features in our data. However we are not completely able to capture the the features towards the final velocities. Our simulation does not produce the higher $\epsilon$ values similar to the experiment (Fig \ref{fig:Simulated} a, inset), or the exact heights of the peaks and valleys at low velocities (Fig \ref{fig:Simulated} b). However most features match, and it is able to generate super-elastic collisions from relatively very simple arguments and parameters. 

\begin{figure}
  
 \includegraphics[width=.5\textwidth]{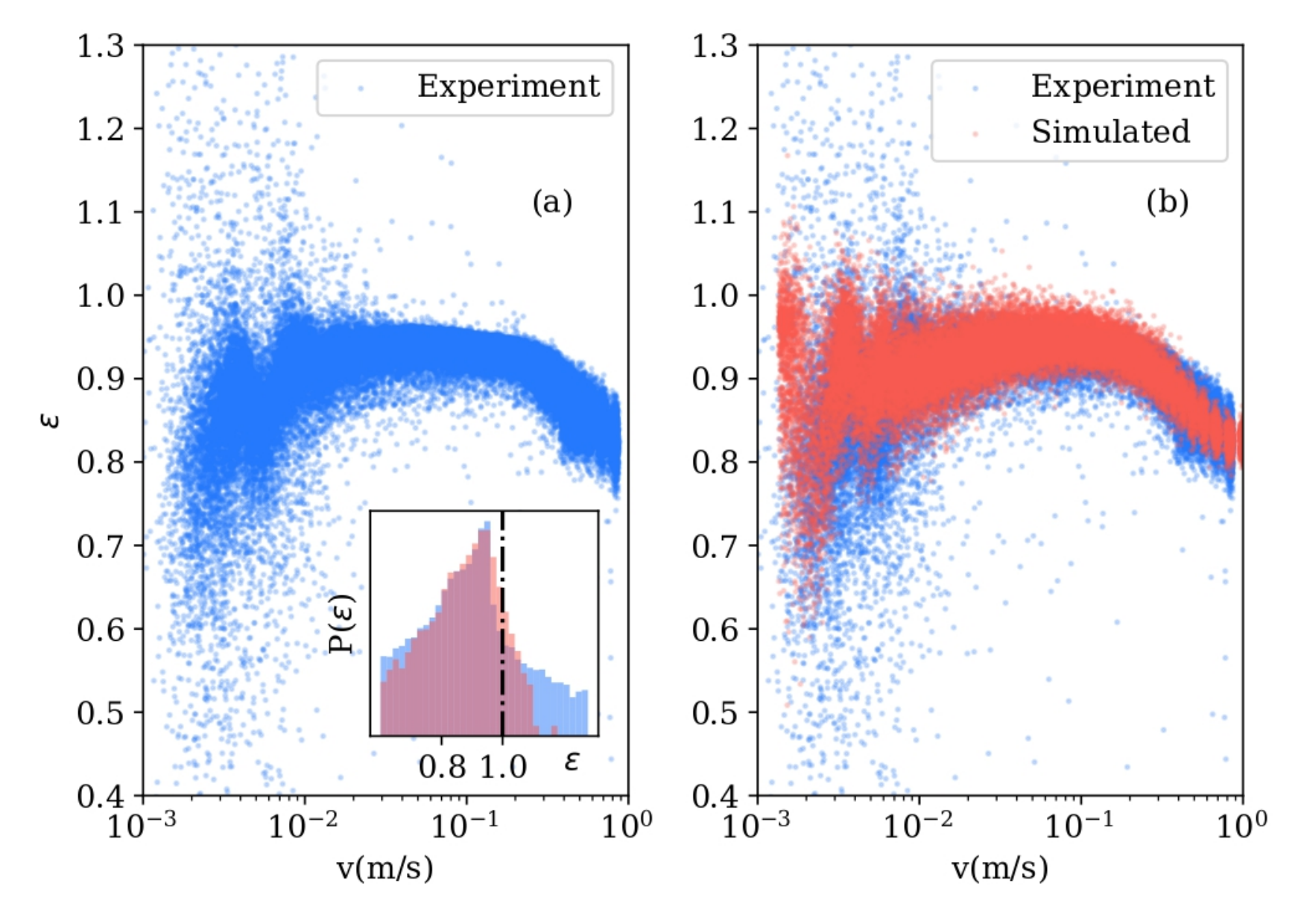}\hfill
  
  \caption{Comparison of simulated and experimental results. Our model fails towards the last bounces, in the final velocities. However we see super-elastic collisions from very simple arguments.  }
  \label{fig:Simulated}
\end{figure}

}
\section{Conclusion}
\rm{We see over the course of our experiments how the bouncing of a steel ball on a steel plate is affected by both the geometry of the plate and the modification of the projectile. Due to the finite size of our target, vibrations set up by the collisions can affect the bouncing, especially at lower velocities, as the vibrations do not damp easily, and persist for multiple collisions. These leads to the last few bounces becoming highly correlated, and the effect of a previous bounce can be seen on a later on in the form of super-elastic collisions, where we believe that the energy imparted to the target from previous collisions constructively add up, and impart into the ball again, leading to $\epsilon$ $>$1. as shown by our simulations.  This argument is further strengthened by the fact that if we add a spring to one end of the ball, we see a sudden rise in the number of super-elastic collisions, although the system as a whole seems to be more lossy than before.\linebreak 
For now our observations are restricted to only a single material.  We would like to further explore this regime, and try to understand if this is simply an odd behaviour we see in our systems or prevalent in all materials.
\section{Acknowledgements}
Funding for this project was made possible due to Tata Institute of Fundamental Research. All materials were machined and built at the Central Workshop of TIFR. We would particularly like to thank Anit Sane, for his invaluable effort in setting up these experiments, and preparing various components. }

\bibliographystyle{ieeetr}


\begin{thebibliography}{10}

\bibitem{PhysRevE.87.042201}
J.~Wakou, H.~Kitagishi, T.~Sakaue, and H.~Nakanishi, ``Inelastic collapse in
  one-dimensional driven systems under gravity,'' {\em Phys. Rev. E}, vol.~87,
  p.~042201, Apr 2013.

\bibitem{RevModPhys.71.435}
L.~P. Kadanoff, ``Built upon sand: Theoretical ideas inspired by granular
  flows,'' {\em Rev. Mod. Phys.}, vol.~71, pp.~435, Jan 1999.

\bibitem{PhysRevE.54.623}
T.~Zhou and L.~P. Kadanoff, ``Inelastic collapse of three particles,'' {\em
  Phys. Rev. E}, vol.~54, pp.~623--628, Jul 1996.

\bibitem{PhysRevE.50.R28}
S.~McNamara and W.~R. Young, ``Inelastic collapse in two dimensions,'' {\em
  Phys. Rev. E}, vol.~50, pp.~R28--R31, Jul 1994.

\bibitem{doi:10.1063/1.858323}
S.~McNamara and W.~R. Young, ``Inelastic collapse and clumping in a
  one‐dimensional granular medium,'' {\em Physics of Fluids A: Fluid
  Dynamics}, vol.~4, no.~3, pp.~496--504, 1992.

\bibitem{delrio2005role}
F.~W. DelRio, M.~P. de~Boer, J.~A. Knapp, E.~D. Reedy, P.~J. Clews, and M.~L.
  Dunn, ``The role of van der waals forces in adhesion of micromachined
  surfaces,'' {\em Nature materials}, vol.~4, no.~8, pp.~629--634, 2005.

\bibitem{maboudian2004surface}
R.~Maboudian and C.~Carraro, ``Surface chemistry and tribology of mems,'' {\em
  Annu. Rev. Phys. Chem.}, vol.~55, pp.~35--54, 2004.

\bibitem{israelachvili1972measurement}
J.~N. Israelachvili and D.~Tabor, ``The measurement of van der waals dispersion
  forces in the range 1.5 to 130 nm,'' {\em Proceedings of the Royal Society of
  London. A. Mathematical and Physical Sciences}, vol.~331, no.~1584,
  pp.~19--38, 1972.

\bibitem{casimir1948influence}
H.~B. Casimir and D.~Polder, ``The influence of retardation on the london-van
  der waals forces,'' {\em Physical Review}, vol.~73, no.~4, p.~360, 1948.

\bibitem{fuller1975effect}
K.~Fuller and D.~Tabor, ``The effect of surface roughness on the adhesion of
  elastic solids,'' {\em Proceedings of the Royal Society of London. A.
  Mathematical and Physical Sciences}, vol.~345, no.~1642, pp.~327--342, 1975.

\bibitem{maugis1996contact}
D.~Maugis, ``On the contact and adhesion of rough surfaces,'' {\em Journal of
  adhesion science and technology}, vol.~10, no.~2, pp.~161--175, 1996.

\bibitem{LandauElasticity}
L.~D. Landau. and E.~M. Lifshitz., {\em Theory of Elasticity}.
\newblock Pergamon, Oxford, 1970.

\bibitem{RamanHS}
C.~V. Raman., ``On some applications of hertz's theory of impact,'' {\em Phys.
  Rev.}, vol.~15, pp.~277--284, Apr 1920.

\bibitem{Zener1941}
C.~Zener, ``The intrinsic inelasticity of large plates,'' {\em Phys. Rev.},
  vol.~59, pp.~669--673, Apr 1941.

\bibitem{Sondergaard1990}
Sondergaard, Chaney, and Brennen, ``Measurements of solid spheres bouncing off
  flat plates,'' {\em Journal of Applied Mechanics}, vol.~112, September 1990.

\bibitem{ktgranular}
N.~V. Brilliantov and T.~Pöschel, {\em Kinetic Theory of Granular Gases}.
\newblock Oxford Graduate Texts.

\bibitem{chastaing2015dynamics}
J.-Y. Chastaing, E.~Bertin, and J.-C. G{\'e}minard, ``Dynamics of a bouncing
  ball,'' {\em American Journal of Physics}, vol.~83, no.~6, pp.~518--524,
  2015.

\bibitem{PhysRevE.48.3988}
J.~M. Luck and A.~Mehta, ``Bouncing ball with a finite restitution: Chattering,
  locking, and chaos,'' {\em Phys. Rev. E}, vol.~48, pp.~3988--3997, Nov 1993.

\bibitem{halev2018bouncing}
A.~Halev and D.~M. Harris, ``Bouncing ball on a vibrating periodic surface,''
  {\em Chaos: An Interdisciplinary Journal of Nonlinear Science}, vol.~28,
  no.~9, p.~096103, 2018.

\bibitem{hsp2}
L.~Labous, A.~D. Rosato, and R.~N. Dave, ``Measurements of collisional
  properties of spheres using high-speed video analysis,'' {\em Phys. Rev. E},
  vol.~56, pp.~5717--5725, Nov 1997.

\bibitem{hsp1}
S.~F. Foerster, M.~Y. Louge, H.~Chang, and K.~Allia, ``Measurements of the
  collision properties of small spheres,'' {\em Physics of Fluids}, vol.~6,
  no.~3, pp.~1108--1115, 1994.

\bibitem{hsp3}
M.~Y. Louge and M.~E. Adams, ``Anomalous behavior of normal kinematic
  restitution in the oblique impacts of a hard sphere on an elastoplastic
  plate,'' {\em Phys. Rev. E}, vol.~65, p.~021303, Jan 2002.

\bibitem{Pendulum_sys1}
F.~G. Bridges, A.~Hatzes, and D.~Lin, ``Structure, stability and evolution of
  saturn's rings,'' {\em Nature}, vol.~309, no.~5966, pp.~333--335, 1984.

\bibitem{Pendulum_sys2}
J.~M. Lifshitz and H.~Kolsky, ``Some experiments on anelastic rebound,'' {\em
  Journal of the Mechanics and Physics of Solids}, vol.~12, no.~1, pp.~35--43,
  1964.

\bibitem{Pendulum_sys3}
G.~Kuwabara and K.~Kono, ``Restitution coefficient in a collision between two
  spheres,'' {\em Japanese Journal of Applied Physics}, vol.~26,
  pp.~1230--1233, aug 1987.

\bibitem{LDV}
W.~Tabakoff, ``Measurements of particles rebound characteristics on materials
  used in gas turbines,'' {\em Journal of Propulsion and Power}, vol.~7, no.~5,
  pp.~805--813, 1991.

\bibitem{PTV}
C.~J. Reagle, J.~M. Delimont, W.~F. Ng, S.~V. Ekkad, and V.~Rajendran,
  ``Measuring the coefficient of restitution of high speed microparticle
  impacts using a ptv and cfd hybrid technique,'' {\em Measurement science and
  technology}, vol.~24, no.~10, p.~105303, 2013.

\bibitem{KINGMENON}
H.~King, R.~White, I.~Maxwell, and N.~Menon, ``Inelastic impact of a sphere on
  a massive plane: Nonmonotonic velocity-dependence of the restitution
  coefficient,'' {\em EPL (Europhysics Letters)}, vol.~93, no.~1, p.~14002,
  2011.

\bibitem{aUDIO}
M.~Heckel, A.~Glielmo, N.~Gunkelmann, and T.~P\"oschel, ``Can we obtain the
  coefficient of restitution from the sound of a bouncing ball?,'' {\em Phys.
  Rev. E}, vol.~93, p.~032901, Mar 2016.

\bibitem{tocvsv}
F.~Gerl and A.~Zippelius, ``Coefficient of restitution for elastic disks,''
  {\em Phys. Rev. E}, vol.~59, pp.~2361, Feb 1999.

\end{thebibliography}
\end{document}